\documentclass[fleqn,10pt]{wlscirep}
\usepackage[utf8]{inputenc}
\usepackage[T1]{fontenc}

\title{A 16 Parts per Trillion Comparison of the Antiproton-to-Proton q/m Ratios}

\author[1,2,3]{M. J. Borchert}
\author[1,4]{J. A. Devlin}
\author[1,4,5]{S. E. Erlewein}
\author[1,6]{M. Fleck}
\author[1,5]{J. A. Harrington}
\author[1,6]{T. Higuchi}
\author[1]{B. Latacz}
\author[1,7]{F. Voelksen}
\author[1,4,5]{E. Wursten}
\author[8]{F. Abbass}
\author[1,5]{M. Bohman}
\author[5]{A. Mooser}
\author[8]{D. Popper}
\author[1,5]{M. Wiesinger}
\author[5]{C. Will}
\author[5]{K. Blaum}
\author[6]{Y. Matsuda}
\author[2,3]{C. Ospelkaus}
\author[7]{W. Quint}
\author[8,9]{J. Walz}
\author[1]{Y. Yamazaki}
\author[1,8]{C. Smorra}
\author[1,*]{S. Ulmer}

\affil[1]{RIKEN, Ulmer Fundamental Symmetries Laboratory, 2-1 Hirosawa, Wako, Saitama, 351-0198, Japan}
\affil[2]{Institut f{\"u}r Quantenoptik, Leibniz Universit{\"a}t Hannover, Welfengarten 1, D-30167 Hannover, Germany}
\affil[3]{Physikalisch-Technische Bundesanstalt, Bundesallee 100, D-38116 Braunschweig, Germany}
\affil[4]{CERN, Esplanade des Particules 1, 1217 Meyrin, Switzerland}
\affil[5]{Max-Planck-Institut f{\"u}r Kernphysik, Saupfercheckweg 1, D-69117, Heidelberg, Germany}
\affil[6]{Graduate School of Arts and Sciences, University of Tokyo, 3-8-1 Komaba, Meguro, Tokyo 153-0041, Japan}
\affil[7]{GSI-Helmholtzzentrum f{\"u}r Schwerionenforschung GmbH, Planckstraße 1, D-64291 Darmstadt, Germany}
\affil[8]{Institut f{\"u}r Physik, Johannes Gutenberg-Universit{\"a}t, Staudinger Weg 7, D-55099 Mainz, Germany}
\affil[9]{Helmholtz-Institut Mainz, Johannes Gutenberg-Universit{\"a}t, Staudingerweg 18, D-55128 Mainz, Germany}
\affil[*]{corresponding author: Stefan Ulmer (stefan.ulmer@cern.ch)}

\begin{abstract}
The Standard Model (SM) of particle physics is both incredibly successful and glaringly incomplete. Among the questions left open is the striking imbalance of matter and antimatter in the observable universe \cite{dine2003origin} which inspires experiments to compare the fundamental properties of matter/antimatter conjugates with high precision \cite{van1987new,ahmadi2018characterization,hori2016buffer, schwingenheuer1995cpt}. Our experiments deal with direct investigations of the fundamental properties of protons and antiprotons, performing spectroscopy in advanced cryogenic Penning-trap systems \cite{Ulmer2015High-precisionRatio}. For instance, we compared the proton/antiproton magnetic moments with 1.5$\,$p$.$p$.$b$.$ fractional precision \cite{smorra2017parts,schneider2017double}, which improved upon previous best measurements \cite{disciacca2013one} by a factor of $>$3000. Here we report on a new comparison of the proton/antiproton charge-to-mass ratios with a fractional uncertainty of 16$\,$p$.$p$.$t$.$ Our result is based on the combination of four independent long term studies, recorded in a total time span of 1.5$\,$years. We use different measurement methods and experimental setups incorporating different systematic effects. The final result,  $-(q/m)_{\mathrm{p}}/(q/m)_{\bar{\mathrm{p}}}$ = $1.000\,000\,000\,003 (16)$, is consistent with the fundamental charge-parity-time (CPT) reversal invariance, and improves the precision of our previous best measurement \cite{Ulmer2015High-precisionRatio} by a factor of 4.3. The measurement tests the SM at an energy scale of $1.96\cdot10^{-27}\,$GeV (C$.$L$.$ 0.68), and improves 10 coefficients of the Standard Model Extension (SME) \cite{ding2020lorentz}. Our cyclotron-clock-study also constrains hypothetical interactions mediating violations of the clock weak equivalence principle  (WEP$_\text{cc}$) for antimatter to a level of $|\alpha_{g}-1| < 1.8 \cdot 10^{-7}$, and enables the first differential test of the WEP$_\text{cc}$ using antiprotons \cite{hughes1991constraints}. From this interpretation we constrain the differential WEP$_\text{cc}$-violating coefficient to $|\alpha_{g,D}-1|<0.030$.
\end{abstract}

\begin{document}

\flushbottom
\maketitle

\thispagestyle{empty}

Various strong motivations to study CPT invariance exist \cite{lehnert2016cpt}; One of them is that CPT symmetry is  inherent to any local, unitary quantum-field-theory without gravity, which is Lorentz invariant, and that has a stable vacuum ground state \cite{luders1957proof}. Tests of CPT invariance therefore constitute probes of the most fundamental pillars of the SM. Furthermore, some approaches to Physics Beyond the SM (PBSM), such as theoretical models with compactified dimensions and non-trivial space-time geometries \cite{edwards2018riemann}, or quantum theories of gravity \cite{tsujikawa2013quintessence,kostelecky1991cpt}, induce CPT-violation. Another motivation to test CPT is that its invariance implies symmetry between the fundamental properties of matter/antimatter conjugates \cite{lehnert2016cpt}, which is in tension with our current best models of the early Universe \cite{weinberg2008cosmology}, predicting  a matter/antimatter balanced radiative universe with a baryon-to-photon ratio of $10^{-18}$. However, cosmological observations indicate a baryon-to-photon ratio of $0.6\times10^{-9}$ \cite{dine2003origin} and a matter-dominated universe, suggesting a possible asymmetry between matter and antimatter. 
\\ 
   \begin{figure*}[htb]
        \centerline
        {
            \includegraphics[width=1\textwidth]
            {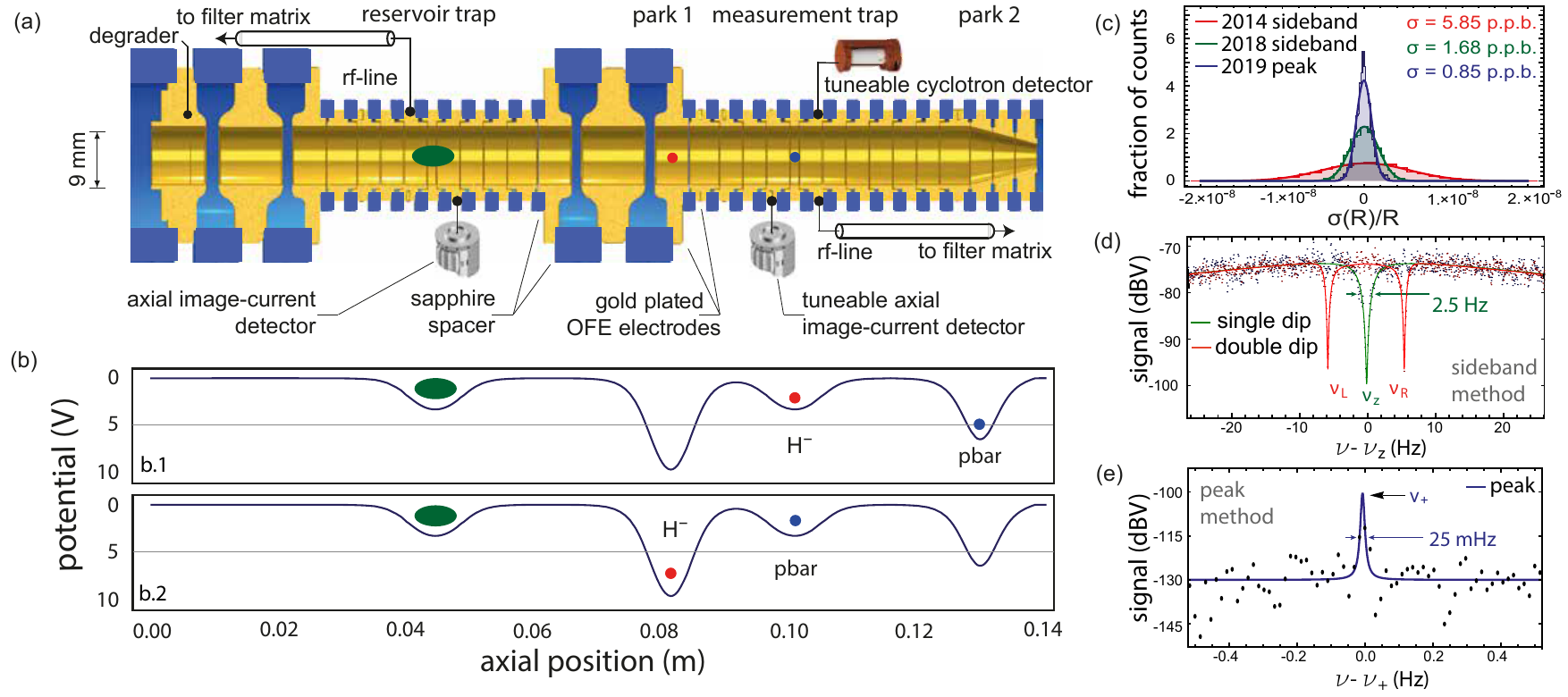}
        }
     \caption{Elements of the experiment to determine the antiproton-to-H$^-$ charge-to-mass ratio. a.) Penning-trap stack, the left trap is a reservoir trap \cite{smorra2015reservoir}, the green ellipse indicates a composite cloud of trapped antiprotons and H$^-$ ions. On the right the measurement trap is shown \cite{Ulmer2015High-precisionRatio}. This trap is equipped with two frequency tuneable detection systems, one for the modified cyclotron frequency $\nu_+$, the other for the axial frequency $\nu_z$.  b.) On axis potential of the trap stack. To compare cyclotron frequencies of H$^-$ ions and antiprotons we shuttle the particles between  configurations b.1 and b.2. c.) Cyclotron frequency ratio fluctuation for the 2014-run (5.85$\,$p.p.b$.$, red \cite{Ulmer2015High-precisionRatio}), the 2019-sideband run (1.67(12)$\,$p.p.b$.$, green) and the 2019-peak run (0.85$\,$p.p.b$.$, blue). The stability improvement between 2014 and 2018 is due to a rigorous redesign of the apparatus \cite{ulmer2019FutureProgram}. d.) Axial dip spectra as used in the sideband method. The dashed green line is a fit to a single particle dip. In case a sideband drive at $\nu_+-\nu_z$ is applied to the trap, the amplitude of the axial oscillator experiences a modulation and a "double-dip" spectrum is observed. The solid red line is a fit to the sideband-spectrum.  e.) Signal of a single antiproton with excited ($E_+=4.5\,$eV) modified cyclotron mode, recorded with the $\nu_+$ image current detector, as used in the peak method. }
     \label{fig:FIG1}
    \end{figure*}

\begin{figure}[htb]
      \centerline{\includegraphics[width=8.5cm,keepaspectratio]{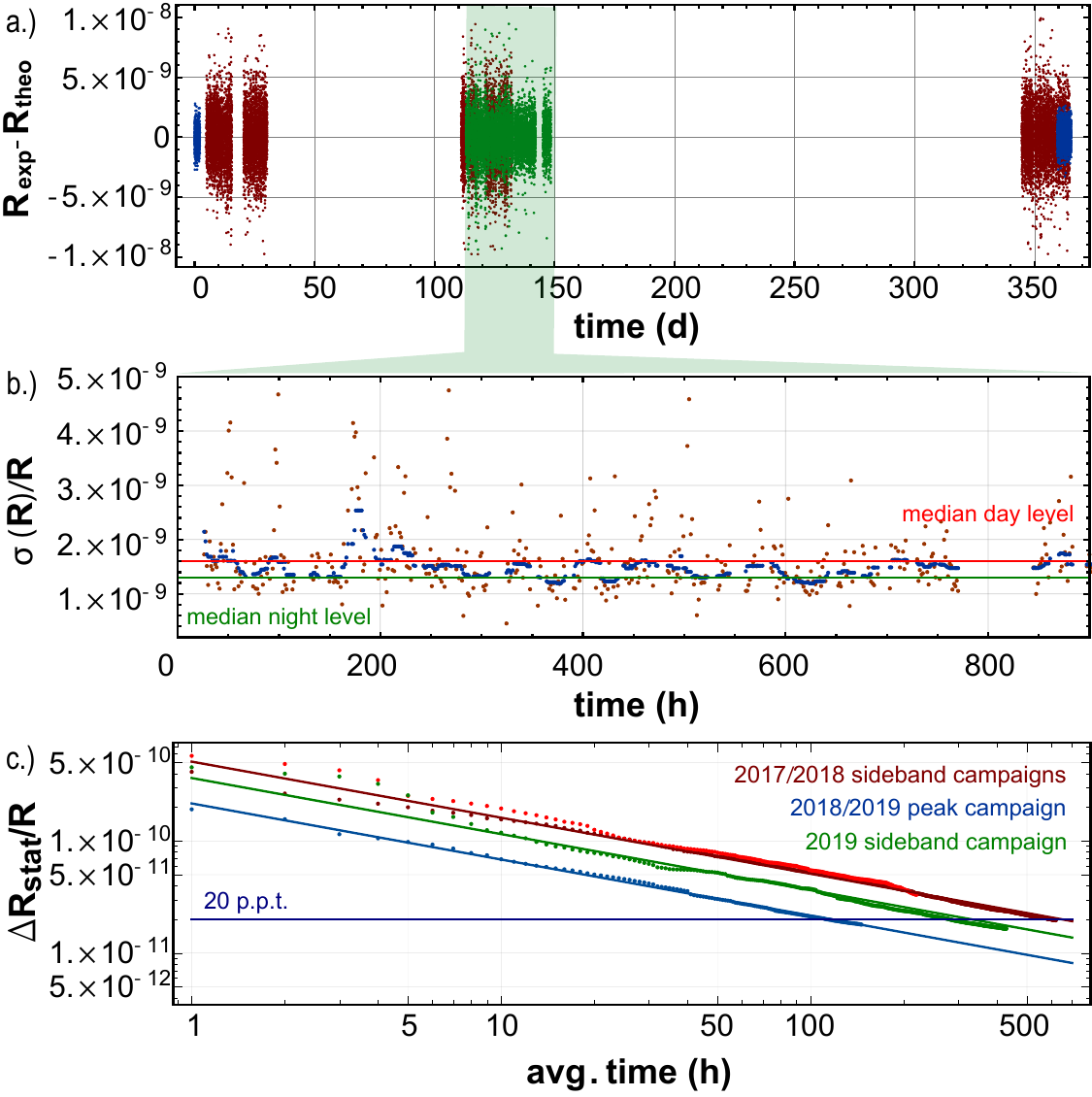}}
     \caption{Results. a.) Recorded data set projected to one sidereal year.  b.) experiment stability $\sigma(R)/R$ as a function of time, each data-point contains and individual data-window of 1.5$\,$h. c.) statistical frequency ratio precision $\Delta R_{\text{stat}}/R$ as a function of averaged measurements. One frequency ratio measurement takes about 260$\,$s. }
     \label{fig:RESULT}
    \end{figure}

For these reasons, direct high-precision experimental tests of CPT invariance provide a topical avenue to search for PBSM. Another fundamental question in physics is whether antimatter obeys the weak equivalence principle (WEP). We study this question by applying the arguments of \cite{hughes1991constraints}, in which anomalous gravitational scalar or tensor couplings to antimatter \cite{hughes1990constraints} would cause clocks formed from matter/antimatter conjugates to oscillate at different frequencies - in direct violation of the WEP$_\text{cc}$. \\
In this article we address both fundamental questions and report on a comparison of the antiproton-to-proton charge-to-mass ratio with a fractional precision of 16$\,$p.p.t. This result improves the precision of our previous best measurement \cite{Ulmer2015High-precisionRatio} by a factor of 4.3 and constitutes the most precise direct test of CPT-invariance with antibaryons. We use the variation of the gravitational potential in our laboratory as the Earth orbits on its elliptical trajectory around the sun to derive stringent limits on scalar and tensor interactions that violate the WEP$_\text{cc}$ for antimatter.  \\   
Our experiment \cite{Smorra2015BASEExperiment} is located at the antiproton decelerator (AD) facility of CERN. It consists of a horizontal superconducting magnet with a homogeneous magnetic field of $B_0 \approx 1.945\,$T that has a temporal stability of $\Delta B_0/B_0\approx 2\,$p$.$p$.$b$.$/h. A cryogenic multi-Penning trap cooled to 4.8$\,$K, see Fig$.\,$\ref{fig:FIG1} (a), is mounted in the center of the magnet bore. The trap is placed inside a vacuum chamber with a volume of 1.2$\,$l. Cryopumping enables lossless antiparticle storage for years \cite{sellner2017improved}, essential for the long term studies reported here. Ultra-stable voltages applied to carefully designed \cite{gabrielse1989open}, gold-plated trap electrodes that are made of oxygen-free electrolytic copper (OFE), provide a locally ideal electrostatic quadrupole potential. 
The trajectory of a single charged particle stored under such electromagnetic conditions can be decomposed into the motion of three independent harmonic oscillators at the modified cyclotron frequency $\nu_+\approx29.6\,$MHz and the magnetron frequency $\nu_-\approx6.9\,$kHz, perpendicular to the magnetic field $B_0\cdot\textbf{e}_z$, and at the axial frequency $\nu_z\approx640\,$kHz, oscillating along the magnetic field lines. The Brown-Gabrielse invariance theorem $\nu_c=(\nu_+^2+\nu_z^2+\nu_-^2)^{1/2}$ relates the three trap frequencies to the free cyclotron frequency $\nu_c=(q B)/(2\pi m)$ \cite{brown1986geonium}. By comparing cyclotron frequencies $\nu_{c,1}$ and $\nu_{c,2}$ of two different particles in the same magnetic field $B_0$, we get access to the ratios of charge-to-mass ratios $\nu_{c,1}/\nu_{c,2}=(q/m)_1/(q/m)_2$.\\
We compare the cyclotron frequencies of single negatively charged hydrogen ions H$^-$ to those of single antiprotons $\bar{p}$ \cite{gabrielse1999precision}. H$^-$ is an excellent negatively charged proxy for the proton (p) with mass 
\begin{eqnarray}
\frac{m_{\text{H}^-}}{m_\text{p}}=1.001\,089\,218\,753\,80(3)\,,
\end{eqnarray}
as detailed in the methods paragraph. 
Comparing particles of the same charge sign avoids inversion of the trapping voltages, and greatly reduces systematic frequency-ratio shifts \cite{Ulmer2015High-precisionRatio}. We measure the individual particle frequencies $\nu_{j,h}$,  $j\in(+,z,-)$ and $h\in (1,2)$, using highly-sensitive superconducting image current detectors \cite{nagahama2016highly}, and apply the particle shuttling method first realized in \cite{Ulmer2015High-precisionRatio}, see Fig$.\,$1 (b). Using this technique, a single frequency ratio comparison takes about $260\,$s. To improve the fractional uncertainty reached in previous experiments \cite{Ulmer2015High-precisionRatio}, numerous experimental upgrades have been implemented.  A rigorous re-design of the cryogenic experiment stage \cite{ulmer2019FutureProgram} and the development of an advanced multi-layer magnetic shielding system \cite{devlin2019superconducting} reduced cyclotron frequency fluctuations by up to a factor of 6, as illustrated in Fig$.\,$\ref{fig:FIG1} (c).  To eliminate the dominant systematic shift of \cite{Ulmer2015High-precisionRatio}, arising from an interplay of trap voltage tuning and residual magnetic field inhomogeneity $B_1$, we have developed a frequency adjustable image-current detector \cite{heisse2017high} for the axial motion oscillating at $\nu_z$. This allows for particle comparisons at constant electrostatic potential and ensures that the antiproton and the H$^-$-ion are compared under exactly the same trapping field conditions. 
\\
To measure the cyclotron frequencies $\nu_{c,\bar{\text{p}}}$ and $\nu_{c,\text{H}^-}$, we prepare the initial conditions shown in Fig$.\,$\ref{fig:FIG1} (b) using the techniques described in \cite{Smorra2015BASEExperiment}. We use two different methods to determine $\nu_+$, one is the well-established sideband-technique (see Fig$.\,$\ref{fig:FIG1} (d)). The other, called the peak-technique, is based on the direct measurement of the modified cyclotron frequency $\nu_+$ \cite{gabrielse1999precision} using a resonant tuneable image-current detector (Fig$.\,$\ref{fig:FIG1} (e)). The sideband-method determines $\nu_+$ by first measuring the axial frequency $\nu_z$. This is accomplished by tuning the particle frequency $\nu_z$ to the detector's resonance frequency $\nu_\text{res}$, and recording a fast Fourier transform (FFT) spectrum of the time transient of the detector output \cite{schneider2017double}. Subsequently, a quadrupolar drive at $\nu_\text{rf}=\nu_+-\nu_z$ is injected to the trap, which leads to an amplitude modulated axial mode oscillation and hence to signal splitting and frequency signatures at $\nu_l$ and $\nu_r$, as shown in  Fig$.\,$\ref{fig:FIG1} (d) \cite{cornell1990mode}. We determine $\nu_z$, $\nu_l$ and $\nu_r$ by least squares fitting to the recorded FFT spectra, and obtain the modified cyclotron frequency as $\nu_+=\nu_\text{rf}+\nu_l+\nu_r-\nu_z$. As all frequency measurements are performed while the particle is in thermal equilibrium with the detection system, the method is largely insensitive to energy-dependent systematic frequency shifts. However, the resolution of the method is intrinsically limited by the 2.5$\,$Hz width of the axial dip and the 25$\,$dB signal-to-noise ratio of the utilized image current detector. At the optimized averaging parameters of the experiment the principal frequency-ratio fluctuation limit of the method is at $\Delta\nu_c/\nu_c=1.67(12)\,$p.p.b. \\
In the peak method the particle's modified cyclotron mode is resonantly excited to energies of order $E_+=4.5\,$eV to $5.5\,$eV, and $\nu_+$ is obtained from a least squares fit to the recorded FFT spectrum shown in Fig$.\,$\ref{fig:FIG1} (e). The observed peak signal has a width about 100 times smaller than the dip signal in the sideband method. 
Since the particle is excited to high $E_+$, the method is however sensitive to energy dependent systematic frequency shifts. Therefore, it is crucial to carefully calibrate $E_{+,\text{H}^-}$ and $E_{+,\bar{\text{p}}}$. Those energies are measured by recording axial frequency shifts dominantly imposed by the residual magnetic inhomogeneity $B_2=-0.0894(6)\,$T/m$^2$ of our measurement trap and relativistic frequency shifts \cite{ketter2014first}. To determine the thermal equilibrium cyclotron frequency $\nu_{+,0}$, we first cool the particle by coupling its modes to the axial detector \cite{cornell1990mode} and measure $\nu_{z,0}$ via the dip method. Afterwards we excite the modified cyclotron mode to an energy $E_+$, and simultaneously record an axial and a peak spectrum to obtain $\nu_{z,\text{exc}}$ and $\nu_{+,\text{exc}}$. Subsequently we determine 
\begin{eqnarray}
\nu_{+,\text{0}}=\nu_{+,\text{exc}}\left(1-\frac{\alpha_+}{\alpha_z} \frac{\nu_{z,\text{exc}}-\nu_{z,0}}{\nu_{z,0}}\right)\,,
\end{eqnarray} 
the trap specific coefficients $\alpha_+$ and $\alpha_z$ are described in the methods paragraph. With this method we achieve a median frequency ratio fluctuation of 850$\,$p.p.t$.$, dominantly limited by magnetic field diffusion.\\ 
The data-set which was recorded, shown in Fig$.\,$\ref{fig:RESULT} a.), consists of $24\,187$ individual frequency ratio measurements acquired within four measurement campaigns between December 2017 and May 2019. The data-set is a clustered Gaussian mixture with superimposed outliers, sourced by changing environmental fluctuations in the accelerator hall, leading to temporal frequency stability fluctuations of the experiment, as shown in Fig$.\,$\ref{fig:RESULT} b.). 
Before analyzing the data, we hence apply robust block stability and median absolute deviation filters \cite{hoaglin2000understanding}, backed-up by magnetometer information. Depending on the run, the filters remove between 1$\,\%$ and 4$\,\%$ of the acquired data, details are described in the supplementary material.    
From the resulting cleaned cyclotron frequency sequences we extract the frequency ratio by superimposing data sets $\nu_{c,\bar{\text{p}},k}$ and $\nu_{c,\text{H}^-,k}$ of length $N_0$ with a free multiplicative estimator $R=\nu_{c,\bar{\text{p}}}/\nu_{c,\text{H}^-}$, while fitting to the resulting data-set a polynomial $p(t)$ of order $q$. We maximise the log-Likelihood function $\log L(\{\nu_{c,\bar{\text{p}},k},\nu_{c,\text{H}^-,k}\},R)$ to find the likeliest $R$-value, and estimate the frequency ratio uncertainty by calculating the Fisher information \cite{le2012asymptotic} $I(R)=d^2\log{L(\{\nu_{c,\bar{\text{p}},k},\nu_{c,\text{H}^-,k}\},R)}/dR^2$ and evaluating the Cramer-Rao lower bound \cite{rao1992information}. Typically, we consider sub-group lengths $N_0$, covering time windows between $1.5\,$h and 4$\,$h, within these time scales other experimental parameters can be considered stable. Depending on the selected sequence length we optimize the order $q$ of the fitting polynomial $p(t)$ using Fisher-ratio tests \cite{Natarayan1993} and reduced Akaike information \cite{wang2006comparison}. To account for the temporal stability variations we evaluate the final frequency ratio  $R_{\bar{\text{p}},\text{H}^-,\text{exp,s}}$ as the weighted arithmetic mean of the determined ratio-sequence $(R,\sigma(R))_k$. The robustness of the data evaluation approach is studied by evaluating the frequency ratio as a function of group-length $N_0$, polynomial order $q$, as well as varying filter-cut conditions. Monte-Carlo simulations relying on a data-based magnetic field model are used to test the evaluation approach. In addition, we benchmark the applied data-evaluation algorithm by comparing identical particles, and obtain
values consistent with 1 within a statistical uncertainty of 14$\,$p.p.t$.$ \\
Processing the acquired data using this evaluation approach and applying systematic corrections summarized in the methods paragraph and the supplementary material, we obtain the results summarized in Tab$.\,$\ref{table:Ratio}.
\begin{table}[htb]
\centering
\begin{tabular}{||c | c | c | c ||} 
\hline
Campaign & $R_\text{exp}$ & $\sigma(R)_\text{stat}$ & $\sigma(R)_\text{sys} $ \\ [0.5ex] 
\hline\hline
2018-1-SB & $1.001\,089\,218\,748$ & $27\cdot10^{-12}$ & $26\cdot10^{-12}$  \\ 
2018-2-SB & $1.001\,089\,218\,727$ & $47\cdot10^{-12}$ & $49\cdot10^{-12}$  \\
2018-3-PK & $1.001\,089\,218\,748$ & $19\cdot10^{-12}$ & $14\cdot10^{-12}$  \\
2019-1-SB & $1.001\,089\,218\,781$ & $19\cdot10^{-12}$ & $23\cdot10^{-12}$  \\
\hline
\end{tabular}
\caption{Results of the four antiproton-to-H$^-$ measurement campaigns. The second column displays the measured result, the last two coulmns indicate the statistical uncertainty and the systematic uncertainty of the measurement.}
\label{table:Ratio}
\end{table}
Figure$\,$\ref{fig:RESULT} c$.$), shows the frequency ratio uncertainty of the different measurement campaigns as a function of averaging time. Between the 2018 sideband runs (red) and the 2019 sideband run (green) the experiment stability was improved by rebuilding the cryogenic support structure of the experiment. The peak method (blue), also performed with the rebuilt instrument, has an intrinsically lower frequency-determination-scatter than the sideband technique.\\
The dominant systematic uncertainty of the sideband campaign arises from a weak scaling of the measured axial frequency $\nu_z$ as a function of its detuning $\Delta\nu_z$ with respect to the resonance frequency $\nu_R$ of the detection resonator. With $\Delta_z=\nu_z-\nu_R$ we determine the function  $\nu_z(\Delta_z)$ based on differential measurements, and extrapolate the result to $\nu_z(\Delta_z=0)$. The leading systematic uncertainty of the peak measurement campaign is due to resolution limits in the determination of the axial temperature of the particles $T_{z,\bar{\text{p}}}$ and $T_{z,{\text{H}^-}}$, respectively. Together with the residual magnetic-bottle inhomogeneity $B_2$ of the trap, a temperature difference $\Delta T_z=T_{z,\bar{\text{p}}}-T_{z,{\text{H}^-}}$ would impose a systematic frequency ratio shift of $\Delta R/R=-70.02\,$p.p.t$./$K and $\Delta R/R=-23.44\,$p.p.t$./$K, for the 2018 sideband runs and the 2018/2019 peak and sideband runs, respectively. For all individual measurement campaigns we determine $\Delta T_z$ using different methods, and correct the measured result accordingly, details are described in the supplementary material. Using a weighted combination of the individual measurement campaigns and accounting for correlations in the systematic uncertainties, we extract 
an antiproton-to-proton charge-to-mass ratio of 
    \begin{eqnarray}
    R_{\bar{\text{p}},\text{p},\text{exp}}=-1.000\,000\,000\,003\, (16)\,.     
    \end{eqnarray}
The result has an experimental uncertainty of 16$\,$p.p.t$.$ (C.L$.$ 0.68), 
supporting CPT invariance. It improves our previous measurement \cite{Ulmer2015High-precisionRatio} by a factor of 4.3 and upon earlier results \cite{gabrielse1999precision} by a factor of 5.6.  \\
In an illustrative model, that can however not be trivially incorporated into relativistic quantum field theory \cite{charlton2020testing}, Hughes and Holzscheiter have shown \cite{hughes1991constraints} that if there was a scalar- or tensor-like gravitational coupling to the energy of antimatter that violates the WEP$_{\text{cc}}$ \cite{hughes1990constraints}, there will be, at the same height in a gravitational field, a frequency difference 
\begin{eqnarray}
\frac{\nu_{c,\bar{\text{p}}}-\nu_{c,{\text{p}}}}{\nu_{c,\text{avg}}}=\frac{3 \Phi}{c^2}\left(\alpha_g-1\right)\,
\end{eqnarray}
between a proton cyclotron-clock at $\nu_{c,{\text{p}}}$ and its CPT conjugate antiproton clock at $\nu_{c,\bar{\text{p}}}$. Here $\alpha_g-1$ is a parameter characterizing the strength of the potential WEP$_\text{cc}$ violation and $\Phi$ the gravitational potential. Together with the gravitational potential of the local supergalactic cluster ($\Phi/c^2=(\text{G M})/(r c^2)=2.99\cdot10^{-5}$) \cite{kenyon1990recalculation,tchernin2020characterizing}, the measurement reported here constrains those WEP$_\text{cc}$ violating gravitational anomalies to a level of $|\alpha_g-1|<1.8\cdot10^{-7}$, improving the previous best limits by about a factor of 4. 
This approach has been discussed controversially \cite{chardin2018gravity}, since the imposed clock shift depends on the absolute value of the gravitational potential, and a WEP-violating force might have a finite range which would modify the chosen potential. 
   \begin{figure}[htb]
      \centerline{\includegraphics[width=8.5cm,keepaspectratio]{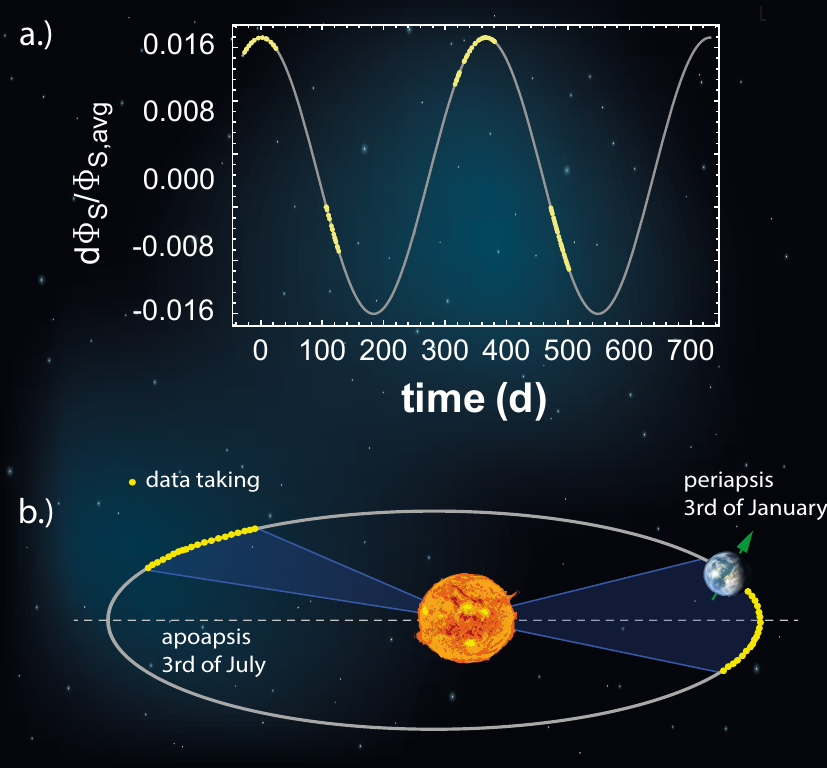}}
     \caption{Trajectory of the earth on its orbit around the sun. a.) Variation of the gravitational potential in the BASE laboratory sourced by the elliptical orbit of the earth around the sun.  The yellow scatter points represent the data-taking windows. b.) Scaled orbit, the blue shaded areas indicate the trajectorial fraction covered by the measurement reported here. }
     \label{fig:Trajectory}
    \end{figure}
This inspires the following analysis: We use the amplitude $d\Phi_\text{S}$ of the change of the mean gravitational potential $\Phi_\text{S,avg}$ at the location of our experiment, which is sourced by the earth's elliptic orbit $O(t)=D_p\cdot(1 - \epsilon^2)/(1 + \epsilon\cos((2\pi/t_\text{sid})t))$ -  with eccentricity $\epsilon=0.017$ and time of the sidereal year $t_\text{sid}$ - 
around the sun. The eccentricity $\epsilon$ leads to a fractional peak-to-peak variation of $2d\Phi_\text{S}/\Phi_\text{S,avg}\approx0.03$, as shown in Fig$.\,$\ref{fig:Trajectory}. In case of WEP violation, this would induce a cyclotron frequency ratio variation  
\begin{eqnarray}
\frac{\Delta R(t)}{R_\text{avg}}=\frac{3 G M_\text{sun}}{c^2}\left(\alpha_{g,D}-1\right)\left(\frac{1}{O(t)}-\frac{1}{O(t_0)}\right)\,,
\end{eqnarray}
where $M_\text{sun}$ is the mass of the sun and $G$ the gravitational constant. 
As shown in Fig$.\,$\ref{fig:Trajectory}, our data set is distributed such that we cover about 80$\,\%$ of the total available peak-to-peak variation of $\Phi_S$. We project our measured frequency ratios to one sidereal year, and look for oscillations of the measured frequency ratio, following the approach described in \cite{abe2017search}. 
From this analysis we derive the differential constraint $|\alpha_{g,D}-1|<0.030$ (C$.$L$.$ 0.68), 
setting limits similar to the initial goals of model-independent experiments testing the weak equivalence principle WEP$_\text{ff}$ by dropping antihydrogen in the gravitational field of the earth \cite{perez2012gbar,bertsche2018prospects,scampoli2014aegis}. \\
At the currently quoted uncertainty of 16$\,$p.p.t$.$, and given that our result is consistent with CPT invariance, this measurement also provides a 4-fold improved limit on the coefficient $r^{H^-}$ of the minimal SME \cite{bluhm1998cpt,kostelecky2011data}, becoming $r^{H^-}<2.09\cdot10^{-27}$ (C.L$.$ 0.68). Very recently an additional non-minimal extension of the SME with up to mass dimension six was applied to Penning-trap experiments comparing particle/antiparticle charge-to-mass ratios \cite{ding2020lorentz}. 
Based on the result presented here, we derive for the charge-to-mass ratio figure of merit defined in \cite{ding2020lorentz}  
\begin{eqnarray}
|\delta\omega_c^{\bar{p}}-R_{\bar{\text{p}},\text{p},\text{exp}}\delta\omega_c^{{p}}-2R_{\bar{\text{p}},\text{p},\text{exp}}\delta\omega_c^{{e}^-}|<1.96\cdot10^{-27}\,\text{GeV}, 
\end{eqnarray}
where the cyclotron frequency differences $\delta\omega_c^{w}$ depend on coefficients $\tilde{b}_w$ and $\tilde{c}_w$, that characterize the strengths of CPT violating background fields, coupling to the involved particles $\bar{\text{p}}$, p, and $\text{e}^-$. In addition to  $r^{H^-}$, our measurement enables us to refine constraints on 9 $\tilde{c}_w$ coefficients of the SME. A time dependent higher harmonic analysis to constrain additional coefficients will be subject of future studies.\\
In conclusion, we have reported on a 4-fold improved measurement of the antiproton-to-proton charge-to-mass ratio with a fractional precision of 16$\,$p.p.t$.$ Our cyclotron frequency comparisons test the SM with an energy resolution of $\approx2\times10^{-27}\,$GeV and improve constraints on CPT violating extensions by a factor of $\approx 4$. Our work enabled us to perform the first differential frequency ratio study with baryonic antimatter to test the WEP$_\text{cc}$ from which we constrain any anomalous gravitational behaviour of antiprotons to $|\alpha_{g,D}-1| < 0.03$. In future measurements, we anticipate to reach even higher sensitivity by improving magnetic field stability and homogeneity, and by the development of transportable antiproton traps, to move precision antiproton experiments from the fluctuating accelerator environment to calm laboratory space, as anticipated by BASE-STEP.

\section*{Methods}

\subsection*{Theoretical Antiproton-to-H$^-$ q/m ratio}
To suppress systematic frequency shifts \cite{Ulmer2015High-precisionRatio}, the measurement presented in the manuscript compares the antiproton to the negatively charged hydrogen (hydride)-ion H$^-$. The use of H$^-$ as a proxy for the proton was first applied in  \cite{gabrielse1999precision}. 
The H$^-$ mass is related to that of the proton by
\begin{eqnarray}
m_{\text{H}^-}=m_p\left(1+2\frac{m_e}{m_p}-\frac{B_e}{m_p}-\frac{A_e}{m_p}+\alpha_{\text{H}^-}\frac{B^2}{m_p}\right),
\end{eqnarray}
where $m_e/m_p$ is the electron-to-proton mass ratio, $B_e/m_p$ is the binding energy of the electron in hydrogen, and $A_e/m_p$ is the affinity energy of the second electron in the electron singlet, both in equivalent proton mass units. The term $\alpha_{\text{H}^-}{B^2}/{m_p}$ is caused by a dynamical frequency shift \cite{thompson2004cyclotron}, related to the electrical polarizability $\alpha_{\text{H}^-}$ of the H$^-$ ion. \\
The leading contribution in Eq$.\,$1 is due to the two additional electrons bound to the H$^-$ ion, and translates to the dominant correction
\begin{eqnarray}
2\frac{m_e}{m_p}=0{.}001\,089\,234\,042\,99(2),
\end{eqnarray}
its uncertainty being about a factor of 1000 below the statistical uncertainty reached in the reported measurement. Here we use the weighted mean of the electron-to-proton mass ratio from Penning trap measurements \cite{heisse2017high, rau2020penning}, and HD$^+$ spectroscopy \cite{kortunov2021proton, patra2020proton}.  \\
For the  binding energy of the electron in hydrogen $B_e/m_p$, we rely on the most recent updates of the NIST atomic spectra database \cite{NIST_ASD}. This value is derived from precision hydrogen spectroscopy results \cite{parthey2011improved} and bound state QED calculations \cite{jentschura2005precise} that contribute to the mass of the hydrogen ion 
\begin{eqnarray}
-\frac{B_e}{m_p}=-0{.}000\,000\,014\,493\,06,
\end{eqnarray}
with uncertainty $<$10$^{-18}$.\\
The best value for the electron affinity energy relies on Doppler-free threshold photodetachment spectroscopy using counter-propagating laser beams performed by Lykke and Lineberger \cite{lykke1991threshold}. They derive an affinity energy of $A_e=-0{.}754\,195(19)\,$eV that contributes 
\begin{eqnarray}
-\frac{A_e}{m_p}=-0{.}000\,000\,000\,803\,81(2).
\end{eqnarray}
The Penning-trap-specific dynamical polarizability shift \cite{thompson2004cyclotron} 
\begin{eqnarray}
\frac{\Delta\nu_c}{\nu_c}=-\alpha_{\text{H}^-}\frac{B^2}{m_p}.
\end{eqnarray}
amounts with the dipole polarizability $\alpha_{\text{H}^-}\approx0{.}3400(8)~\cdot~10^{-38}\,\text{C} \text{m}^2/\text{V}$ \cite{sahoo2020determination} of the H$^-$ ion and the BASE magnetic field of 
$B_0\approx1{.}944\,8\,$T, to
\begin{eqnarray}
\alpha_{\text{H}^-}\frac{B^2}{m_p}=7{.}689(18)\,\text{p.p.t}\,. 
\end{eqnarray}
Taking all the corrections into account, the theoretical proton-to-H$^-$ cyclotron frequency ratio is given as 
\begin{eqnarray}
R_{p,{\text{H}^-},\text{theo}}=1{.}001\,089\,218\,753\,80(3)\,.
\end{eqnarray}

\subsection*{Sideband Method and Limits}
Sideband measurement methods \cite{cornell1990mode} rely on the determination of the modified cyclotron frequency $\nu_{+}$, that are  entirely based on thermal equilibrium measurements. We first record a single particle dip spectrum (56$\,$s) and obtain the axial frequency $\nu_z$ by performing a least squares fit to the recorded spectrum. Afterwards, a radio-frequency drive at the sideband frequency $\nu_\text{rf}\approx\nu_+-\nu_z$ (sideband drive) is applied, while simultaneously the noise transient of the axial detector's output is recorded ($64\,$s) and an FFT is performed on those data. This results in a {double dip spectrum}, from which the frequencies $\nu_l$ and $\nu_r$ are extracted, also based on least squares fitting. Based on these frequency measurements, the modified cyclotron frequency is determined by 
\begin{eqnarray}
\nu_+=\nu_\text{rf}+\nu_l+\nu_r-\nu_z.
\end{eqnarray}
Together with the measurement of the axial frequency $\nu_z$, the magnetron frequency is derived as $\nu_{-}\approx\nu_z^2/(2\nu_+)$ \cite{brown1986geonium}. Application of the invariance theorem \cite{brown1986geonium} 
    \begin{eqnarray}
    \nu_{c}&=& \sqrt{\nu_+^2+\nu_z^2+\nu_-^2}
    \label{eq:invariance}
    \end{eqnarray}
yields the free cyclotron frequency $\nu_c$. \\
The cyclotron frequency ratio scatter in sideband measurements is determined by the axial frequency fluctuation $\sigma(\nu_z)$ of the fit of the dip-lineshape to the recorded spectrum, which is
\begin{eqnarray}
\sigma(\nu_z)=\frac{0.443(18)}{\sqrt{t_\text{avg}}}+ 3.7(5.2)\cdot10^{-6} t_\text{avg}\, (\text{Hz}).
\end{eqnarray}
Given this stability, the resulting frequency ratio scatter, assuming constant magnetic field, is expected to be 
\begin{eqnarray}
\sigma(R)&\approx&\sqrt{\Xi_z^2+2\left(\sqrt{\frac{\Delta\nu_{z,\text{SB}}}{\Delta\nu_{z}}}\frac{\text{SNR}_{z}}{\text{SNR}_{z,\text{SB}}}\Xi_z\right)^2}\\
&=&1.67(12)\,\text{p.p.b.}
\end{eqnarray}
where $\Xi_z$ is the background fluctuation of the axial frequency, $\Delta\nu_{z}$ and $\Delta\nu_{z,\text{SB}}$ are the widths of the single dip and the double dip, and SNR$_{z}$ and SNR$_{z,\text{SB}}$ their signal-to-noise ratios, respectively. In the approximation we neglect common mode noise by the power supply which is at a level of $\approx 14\,$mHz for the frequency averaging times considered here.

\subsection*{Peak Method and Limits}
\begin{figure}[h!]
      \centerline{\includegraphics[width=8.5cm,keepaspectratio]{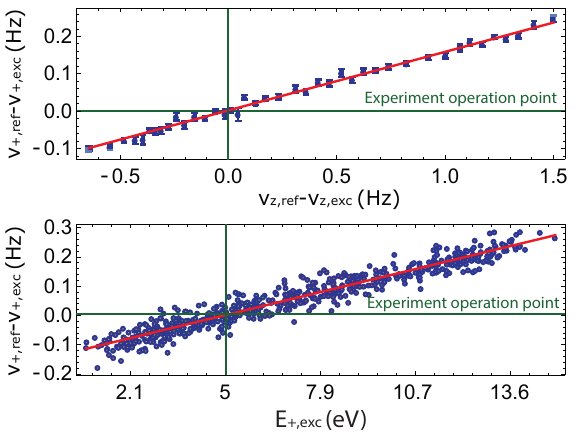}}
     \caption{Upper: Measured cyclotron frequency shift as a function of the measured axial frequency shift. Lower: Measured cyclotron frequency shift as a function of partcile energy $E_+$.  }
     \label{fig:PLOT_Linear}
    \end{figure}

In the peak method, a modified version of the technique described in \cite{gabrielse1999precision}, we determine the modified cyclotron frequency $\nu_{+,0}$ by direct observation of an excited single trapped particle, using the dedicated cyclotron image current detection system. To perform a single modified cyclotron frequency measurement we execute the following sequence:
\begin{enumerate}
    \item We tune the axial resonator to resonance with the axial frequency of the particle of interest.
    \item We cool the modified cyclotron mode $\nu_{+}$ by applying a sideband drive \cite{cornell1990mode} at $\nu_{+}-\nu_{z}$. This drive is typically applied for a few seconds and defines the initial thermal energy spread $E_\text{th}$ of the particle.
    \item We measure the axial frequency $\nu_{z,\text{ref}}$ of the particle in thermal equilibrium with the detection system, this reference axial frequency measurement typically takes 42$\,$s.  
    \item We excite the particle's modified cyclotron mode using a bursted resonant drive at $\nu_+$. The drive is chosen such that the particle is typically excited to a mode energy of  $E_+=4.5\,$eV to $E_+=5.5\,$eV, corresponding to a particle radius of $150\,\mu$m to $200\,\mu$m, which produces on the FFT of the recorded time transient a peak-signal with a signal-to-noise ratio of typically $17.5(2.5)\,$dB ($\approx 7.5$ in linear units) and a full width at half maximum of 27.5(3.5\,)mHz. For the excitation, the amplitude and burst parameters are chosen such that the excitation drive typically interacts with the particle for about 700$\,\mu$s.
    \item Afterwards the actual frequency measurement takes place, which simultaneously records the modified cyclotron frequency $\nu_{+,\text{exc}}(E_+)$ and the axial frequency of the excited particle $\nu_{z,\text{exc}}(E_+)$ for about $t_\text{exp}=66\,$s. Note that the measurement-to-cooling-time ratio is $t_\text{exp}/\tau_+\approx0.15$. 
\end{enumerate}
From measurements $5.$ and $3.$ we obtain $\nu_{+,\text{exc}}(E_+)$, $\nu_{z,\text{exc}}(E_+)$ and $\nu_{z,\text{ref}}(E_+=E_\text{th})$, respectively. Assuming that frequency shifts are linear in $E_+$, we obtain 
\begin{eqnarray}
\label{Eq:1}
\nu_{+,\text{exc}}=\nu_{+,0,\text{exp}}\left(1+\alpha_+ E_{+}\frac{\tau_+}{t_\text{exp}}\left(1-\exp{(-t_\text{exp}/\tau_+)}\right)\right)\,,
\end{eqnarray}
and
\begin{eqnarray}
\label{Eq:2}
\nu_{z,\text{exc}}=\nu_{z,0,\text{exp}}\left(1+\alpha_z E_{+}\frac{\tau_+}{t_\text{exp}}\left(1-\exp{(-t_\text{exp}/\tau_+)}\right)\right)\,,
\end{eqnarray}
where $\alpha_+$ and $\alpha_z$ are trap specific coefficients. This measurement concept enables us to solve 
\begin{eqnarray}
\label{Eq:3}
E_{+}\frac{\tau_+}{t_\text{exp}}\left(1-\exp{(-t_\text{exp}/\tau_+)}\right)={\alpha_z}\frac{\nu_{z,\text{exc}}-\nu_{z,0,\text{exp}}}{\nu_{z,\text{ref}}}\,,
\end{eqnarray}
and to rewrite 
\begin{eqnarray}
\label{Eq:5}
\nu_{+,0,\text{exp}}=\nu_{+,\text{exc}}\left(1-\frac{\alpha_+}{\alpha_z} \frac{\nu_{z,\text{exc}}-\nu_{z,\text{ref}}}{\nu_{z,\text{ref}}}\right),
\end{eqnarray}
which allows us to extract the frequency of interest $\nu_{+,0,\text{exp}}$. \\
To dominant order and for the trap which is used in the experiment, the coefficients $\alpha_+$ and $\alpha_z$ are defined by residual magnetic inhomogeneity, relativistic effects and trap anharmonicities. The cyclotron coefficient reads 
\begin{eqnarray}
\alpha_+&=&-\frac{1}{m c^2}\nonumber\\
&-&\frac{1}{4\pi^2m \nu_{z}^2}\left(\left(\frac{B_1}{B_0}\right)^2+\left(\frac{\nu_{z}}{\nu_{+}}\right)^2\left(\frac{B_2}{B_0}\right)\right)\nonumber\\
&+&\frac{3}{4}\frac{1}{qV_0}\left( \frac{\nu_z}{\nu_+}\right)^4\left(\frac{C_4}{C_2^2}\right)
\end{eqnarray}
where the first term is the relativistic shift. The terms in the second row of the equation are sourced by magnetic field inhomogeneities. The first arises from the force $F=-\mu_+(dB/dz)$ which counteracts the trapping potential and shifts the particle along the trap axis, here $\mu_+$ is the angular magnetic moment associated with the trajectory of the modified cyclotron motion, the $B_2$ term is purely geometric. The term in the third row arises from the octupolar contribution $C_4$ of the electrostatic trapping potential which modifies, compared to a purely quadrupolar potential $V_0 C_2(z^2-\rho^2/2)$, the strength of the radially pulling electrostatic force.\\ 
The axial coefficient reads   
\begin{eqnarray}
\alpha_z&=&-\frac{1}{2m c^2}\nonumber\\
&+&\frac{1}{4\pi^2m \nu_{z}^2}\frac{B_2}{B_0}\nonumber\\
&-& \frac{3}{2}\frac{1}{qV_0}\left(\frac{\nu_z}{\nu_+}\right)^2\left(\frac{C_4}{C_2^2}\right),
\end{eqnarray}
the first term is relativistic. Similar to the continuous Stern-Gerlach Effect, the source of the second term is the interaction of $\mu_+$ with the residual magnetic bottle $B_2$ of the trap. The third term arises from the octupolar component of the trapping potential which modifies the potential curvature experienced by the axial oscillator.  
The coefficient $\alpha_+$ is dominated by the relativistic shift $\Delta\nu_+/\nu_+\approx -1\,$p.p.b$.$/eV$\cdot E_+$. The contributions of the magnetic bottle and the magnetic gradient term are suppressed by a factor of $\approx8.4$ and $\approx280$, respectively. The coefficient $\alpha_z$ is mainly defined by the magnetic bottle term $B_2$, compared to this dominant effect the relativistic shift is suppressed by a factor of $\approx490$. During the experiment campaign we measure the coefficient $B_2$ once in 4.5$\,$h. The coefficient $B_1$ leads to a marginal frequency shift and is therefore only determined at the beginning and the end of a measurement campaign. \\
The coefficient $C_4$ is a trap tuning coefficient which depends on the "tuning ratio" $\text{TR}=V_\text{CE}/V_0$, where $V_\text{CE}$ is the voltage applied to the correction electrodes and $V_0$ the voltage applied to the central ring electrode of the trap. The $C_4$ coefficient contributes an axial frequency shift of $\Delta\nu_z=367.21\,$mHz/(mUnit$\cdot$eV)$\cdot\Delta\text{TR}\cdot E_+$ and a modified cyclotron frequency shift of $\Delta\nu_+=3.97\,$mHz/(mUnit$\cdot$eV)$\cdot\Delta\text{TR}\cdot E_+$. Here, $\Delta\text{TR}$ is the shift from the optimum tuning ratio defined by $C_4(\text{TR})=0$ and the practically applied experimental tuning ratio. We regularly optimize the TR, given the properties of our power-supplies, $\Delta\text{TR}$ can be optimized to a level of $6\cdot10^{-6}$ and is stable within this level for months. The resulting residual uncertainty in $C_4$ leads to uncertainties in the axial  and the modified cyclotron frequency shift of $\Delta\nu_z=2.20\,$mHz/eV$\cdot E_+$ and  $\Delta\nu_+=23.82\,\mu$Hz/eV$\cdot E_+$, respectively. \\
To test the experimental principle, we measure the cyclotron frequency as a function of the axial frequency for different excitation energies $E_+$, results are shown in Fig$.\,$\ref{fig:PLOT_Linear}. We measure frequency differences $\nu_{+,\text{ref}}-\nu_{+,\text{exc}}$ and $\nu_{z,\text{ref}}-\nu_{z,\text{exc}}$ as a function of particle excitation energy, as shown in the upper graph. Using the measured $B_2$ coefficient, in the lower plot the axial frequency difference is scaled to excitation energy. Within an excitation energy-span of about 12$\,$eV we observe a clear linear scaling, the experiment is operated at a median energy of $5.03\,$eV, indicated by the green lines. 
The impact of residual nonlinear contributions and uncertainties $\Delta_+$ and $\Delta_z$ in the coefficients $\alpha_+$ and $\alpha_z$ are discussed in the supplementary material and below. \\
The observed median frequency ratio scatter of the peak measurement campaign is $\sigma(R)\approx850\,$p.p.t$.$, limited by the current stability of the superconducting magnet. The principal resolution limit of the method is constituted by an interplay of energy determination- and cyclotron frequency determination scatter. For our detector parameters, excitation energies, and the chosen averaging times, the principal limit is at $\sigma(R)\approx380\,$p.p.t$.$ while the best scatter observed in the experiment was at $\sigma(R)\approx500\,$p.p.t$.$, the discrepancy contributed by magnetic field fluctuations.

\subsection*{Sideband Method: Dip-Line-shape}
The recorded axial frequency dip spectra are fitted by the parallel tuned circuit lineshape model described in \cite{wineland1975principles}
\begin{eqnarray}
\chi(\nu)=\frac{R_p(\nu^2-\nu_p^2)^2}{(\nu^2-\nu_p^2)^2+(\frac{Q}{\nu_R\nu}(\nu^2-\nu_p^2)(\nu^2-\nu_R^2)+\Delta\nu_z\nu)^2},
\label{Eq:Lineshape}
\end{eqnarray}
where $R_p=2\pi\nu_R Q L$ is the resonant effective parallel resistance of the detector, $Q$ and $\nu_R$ the quality factor and resonance frequency of the detector, respectively, and $L$ the inductance of the detection toroid. The particle/detector interaction damps the particle's motion and induces a dip line width $\Delta\nu_z=1/(2\pi\tau)$, $\tau = (m/R_p)(D_\text{eff}/q)^2$ being the resistive damping time constant and $D_\text{eff}=10.02\,$mm the trap specific pickup length. 
In the experiment we measure the thermal noise power of the axial detector's output which we parameterize as 
\begin{equation}
\begin{split}
S=F(\nu)\left(10\cdot\log\left(4k_BT_z\chi(\nu)\kappa^2+e_n(\nu)^2\right)\cdot G(\nu)\right)\nonumber\\
+ F(\nu)\left(\sum_k A_k(\nu-\nu_R)^k\right) ,
\end{split}
\end{equation}
where $\kappa$ \cite{nagahama2016highly} defines the amplifier-to-resonator coupling, $e_n (\nu)$ is the equivalent input noise of the low-noise amplifier connected to the detection resonator, $G(\nu)$ is the gain function of the detector, $A_k$ are phenomenological shape coefficients and $F(\nu)$ describes the input characteristic of the FFT-analyzer. Practically, we record in each measurement a broadband FFT spectrum of the resonator (400$\,$Hz) as well as a narrow-band spectrum (50$\,$Hz). We fit the undisturbed lineshape to the measured narrow-band spectra, use the broadband spectra to determine deviations from the ideal lineshape-model, and perform, given the characterized deviations, perturbation theory on frequency shifts imposed by hidden effects consistent with the power of the fit residuals. Our analysis includes the determination of resonator shape coefficients and effects arising from FFT distortions, summarized in Tab$.\,$\ref{table:SUMM}. Effects related to 1/f-amplifier noise and frequency scaling of amplifier gain and FFT input characteristics contribute $<0.01\,$p.p.t$.$ and are not listed explicitly.  

\subsection*{Sideband Method - Spectrum Shift }
\begin{figure}[h!]
      \centerline{\includegraphics[width=8.5cm,keepaspectratio]{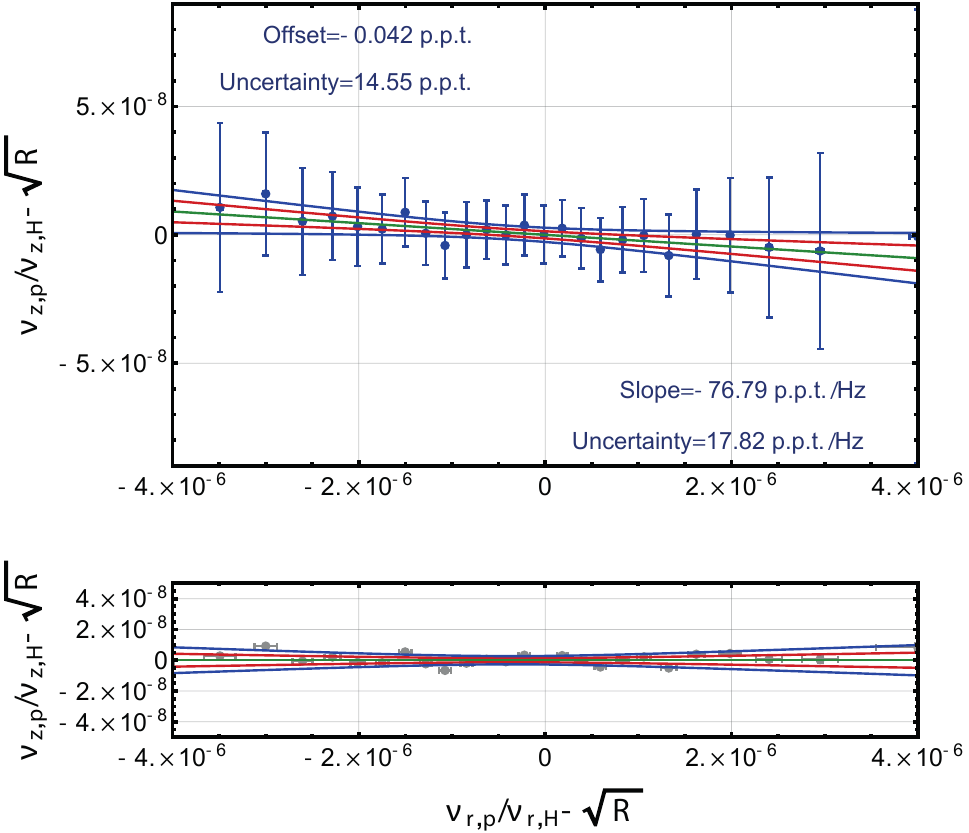}}
     \caption{Upper: measured axial frequency ratio as a function of the frequency ratio of the axial detection resonators. We observe a weak linear scaling of the measured axial frequency ratio as a function of the detuning of the axial frequency with respect to the resonator center. Green line: weighted linear fit, red and blue functions represent C$.$L$.$0.68 and C$.$L$.$0.95 error bands. Lower: Residuals of upper plot.   }
     \label{fig:PLOT_Dominant}
    \end{figure}

For all measurement campaigns we observe a linear scaling of the measured axial frequency $\nu_z$ as a function of the position of $\nu_z$ with respect to the resonance frequency $\nu_R$ of the detection resonator. 
While this shift is negligibly small when using the peak technique, it is of dominant concern in sideband measurements, since any shift in the measured axial frequency translates directly into a shift of the obtained modified cyclotron frequency. To quantify this effect we compare interleaved antiproton/antiproton, H$^-$/H$^-$, and antiproton/H$^-$ axial frequency measurements and evaluate $\nu_{z,j,\text{det}}/\nu_{z,j}-\sqrt{R_{p,{\text{H}^-},\text{theo}}}$ as a function of the fractional resonator/resonator frequency ratio $\nu_{R,j,\text{det}}/\nu_{R,j}-\sqrt{R_{p,{\text{H}^-},\text{theo}}}$. An example of such a measured scaling is shown in  Fig$.\,$\ref{fig:PLOT_Dominant}, where antiproton and H$^-$ frequencies are compared.  
To study how this effect affects the left and right frequency components of a double dip measurement, we investigate the scaling of the quantity $\nu_l+\nu_r-\nu_z$ as a function of $\nu_z-\nu_R$ and determine the linear slope $m$ of the resulting data using weighted linear fits. Throughout the experiment sequence and imposed by slow drifts of the resonator frequency and trapping voltages, integrated residual shifts $\nu_{z,\bar{\text{p}}}-\nu_{R,\bar{\text{p}}}$ and $\nu_{z,{\text{H}^-}}-\nu_{R,{\text{H}^-}}$ accumulate, which impose a potential systematic shift to the determined cyclotron frequency ratios. 
We correct for each data sub-set the median frequency difference accumulated over the respective sequence by the slope $m$, which projects the frequency ratio to  $(\nu_{z,\bar{\text{p}}}-\nu_{R,\bar{\text{p}}})-(\nu_{z,{\text{H}^-}}-\nu_{R,{\text{H}^-}})=0$. Table \ref{table:CORR} summarizes all the corrections applied to the available sub-datasets, for all measurements the corrections are within the resolution consistent with zero and shift the result by less than 30$\,\%$ of the total uncertainty quoted for the respective measurement. Given the available statistical resolution of the frequency scaling as a function of detuning $\nu_z-\nu_R$, this line-shape correction contributes the dominant systematic uncertainty of the sideband campaigns. For the peak measurements the related systematic shift is suppressed by $(\nu_z/\nu_+)$.   

\subsection*{Voltage Drifts}
During the measurement, the particles are transported from the upstream and the downstream park electrodes into the measurement trap. Different relaxation times of the different voltage supply channels, as well as different time constants of the filter electrodes can therefore lead to systematic axial frequency shifts, which is of concern for the sideband method measurements. To characterize these shifts we measure axial frequencies of particles transported from the upstream and downstream electrodes into the measurement trap. We obtain the drift offset as 
\begin{eqnarray}
\Delta_{z,D}=\frac{1}{2}\left(\Delta_{u,d}+\Delta_{d,u}\right)
\end{eqnarray}
where $\Delta_{u,d}=\nu_{z,\bar{\text{p}},u}-\nu_{z,\text{H}^-,d}$ is the axial frequency difference with the antiproton transported into the trap from the upstream side and H$^-$ from downstream, while $\Delta_{d,u}=\nu_{z,\text{H}^-,u}-\nu_{z,\bar{\text{p}},d}$ is the frequency for interchanged particles. We combine these results with explicit identical particle measurements and obtain $\Delta\nu_{z,D}=0.149(152)\,$mHz, within the resolution of the measurement consistent with 0. During the first axial frequency measurement after particle transport the downstream particle appears to have an axial frequency which is slightly shifted upwards compared to the particle in the upstream electrode. We consider this shift for the individual particle/electrode configurations of the sideband measurement campaigns, in the peak campaign the effect is suppressed by $\nu_z/\nu_c$ and therefore negligibly small. Magnetic field shifts imposed by this residual drift can be constrained to the sub 0.1$\,$p.p.t.-level. 

\subsection*{Peak Method - First Order Coefficient Shifts}
\begin{figure}[h!]
      \centerline{\includegraphics[width=7.0cm,keepaspectratio]{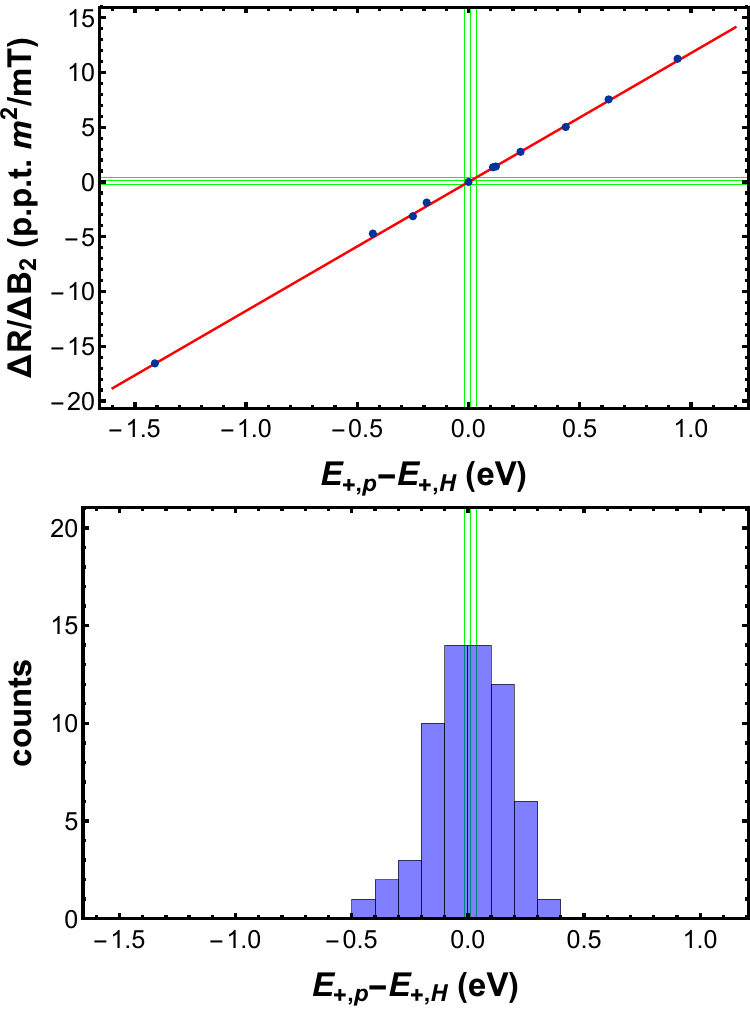}}
     \caption{Upper: Sensitivity of the frequency ratio $R$ as a function of the coefficient $B_2$ for different particle energy differences $E_{+,\text{p}}-E_{+,\text{H}}$, expressed as $\Delta R/\Delta B_2 (\Delta E_+)$ Lower: Measured particle energy differences $E_{+,\text{p}}-E_{+,\text{H}}$ throughout the peak run. The green vertical lines indicate the mean difference and the uncertainty, the vertical green lines define the frequency ratio shift and its uncertainty caused by the uncertainties in energy similarity and $B_2$.}
     \label{fig:dRB2}
 \end{figure}

In this section we study systematic frequency ratio shifts that are imposed by the experimental uncertainties and shifts $\Delta_+$ and $\Delta_z$ in the experimental coefficients $\alpha_+$ and $\alpha_z$, by cyclotron resonator cooling-time constant differences $\tau_{+,\bar{\text{p}}}\neq\tau_{+,\text{H}^-}$, and differences in the excitation energies $E_{+,\bar{\text{p}}}\neq E_{+,\text{H}^-}$. The fractional frequency shift $\Delta\nu_+/\nu_+$ between the experimentally determined cyclotron frequency $\nu_{+,0,\text{exp}}$ and the real cyclotron frequency $\nu_{+,0}$ is given as
\begin{eqnarray}
\frac{\Delta\nu_+}{\nu_{+}}&=&\left(\Delta_+-\frac{\alpha_{+}}{\alpha_z}\Delta_z\right)\nonumber\\
&\cdot&\left(1-\frac{1}{2}\frac{t_{\text{avg}}}{\tau_{+,\bar{\text{p}}}}+\frac{1}{6}\left(\frac{t_{\text{avg}}}{\tau_{+,\bar{\text{p}}}}\right)^2\right)E_{+}
\end{eqnarray}
Incorrectly assigned coefficients $\alpha_{k,e}=\alpha_k+\Delta_k$ lead to systematic frequency ratio shifts, especially significant, once the particles are excited to different energies $E_{+,\bar{\text{p}}}$ and $E_{+,{\text{H}^-}}$, respectively. In this case, the resulting first-order frequency ratio shift reads 
\begin{eqnarray}
\frac{\Delta R}{R}=\left(\Delta_+-\frac{\alpha_{+}}{\alpha_z}\Delta_z\right)\left(E_{+,\bar{\text{p}}}-E_{+,{\text{H}^-}}\right)\,,
\end{eqnarray}
differences $\Delta\tau_+$ in the cooling time constants $\tau_{+,\bar{\text{p}}}$ and $\tau_{+,{\text{H}^-}}$ are suppressed as $({t_{+,\text{avg}}})/(2{\tau_+})({\Delta_+}/{\tau_+})$.
To characterize frequency ratio shifts imposed by $\Delta_k$,  we measure the cyclotron frequency ratio at deliberately different particle energies $E_{+,\bar{\text{p}}}$ and $E_{+,{\text{H}^-}}$. Subsequently we evaluate $R$ as a function of $\alpha_+(B_k,C_k)$ and $\alpha_z(B_k,C_k)$ - $B_k$ and $C_k$ characterizing trap anharmonicities - and determine the slopes $dR/dB_k$ and $dR/dC_k$, 
an example result is shown in Fig$.\,$\ref{fig:dRB2}. The upper graph displays the derivative of the frequency ratio shift $\Delta R$ with respect to a shift of the dominant correction coefficient $\Delta B_2$, as a function of particle energy difference $E_{+,\bar{\text{p}}}-E_{+,{\text{H}^-}}$. The lower plot shows a histogram of the particle energy differences $E_{+,\bar{\text{p}}}-E_{+,{\text{H}^-}}$ of the considered peak-ratio data-set. The determined slope $d(\Delta R/\Delta B_2)/d(\Delta E_+)=11.77(10)$p.p.t$.\cdot$m$^2$/mT, the uncertainty on energy similarity $\sigma(E_{+,\bar{\text{p}}}-E_{+,{\text{H}^-}})=0.009(26)\,$eV, and the uncertainty of the experimental coefficients $\alpha_{+}$ and $\alpha_z$ lead to a shift of the measured frequency ratio of $\Delta R=0.16(40)\,$p.p.t$.$, which needs to be corrected.

\subsection*{Dominant Systematic Trap Shift}
The dominant trap related systematic frequency shift arises from an interplay of the weakly bound axial oscillator with the residual magnetic bottle $B_2$ of the trap. In the residual $B_2$ inhomogeneity the axial oscillator, which is in contact with the axial thermal reservoir, averages as a function of axial energy over a mean magnetic field. This induces a fractional cyclotron frequency shift 
\begin{eqnarray}
\frac{\Delta\nu_+}{\nu_+}=\frac{1}{4\pi^2 m \nu_z^2}\frac{B_2}{B_0}k_BT_z,
\end{eqnarray}
where $k_B$ is the Boltzmann constant and $T_z$ the temperature of the axial resonator. With the parameters of our experiment this induces a fractional frequency shift of $262.24\,$p.p.t$.$ m$^2$/(T$\cdot$K)$\cdot B_2 T_z$, for the 2018 sideband-run with $B_2=-0.267(2)\,$T/m$^2$ and the 2018 peak- and 2019 sideband-runs with $B_2=-0.0894(6)\,$T/m$^2$ the imposed shifts are $-70.02\,$p.p.t$.$/K$\cdot T_z$ and $-23.44\,$p.p.t$.$/K$\cdot T_z$, respectively. \\
In addition, in both applied measurement sequences the particles are sideband-cooled by coupling $\nu_+$ to the axial resonator. This induces a relativistic shift of the measured cyclotron frequency ratio
\begin{eqnarray}
\Delta R\approx\frac{\nu_{c,\bar{\text{p}}}}{\nu_{c,\text{H}^-}}=\frac{1}{mc^2}\frac{\nu_+}{\nu_z}k_B\Delta T_z
\end{eqnarray}
which is at a level of $\Delta R=4.25\,$p.p.t$.$/K$\cdot\Delta T_z$. \\
Any axial temperature difference $\Delta T_z$ between the antiproton and the H$^-$ ion would therefore induce considerable systematic frequency ratio shifts, which requires careful axial temperature comparisons of the particles.\\   
To determine the axial temperature $T_z$, we use a combination of axial frequency measurements, sideband-cooling drives, and resonant excitation of the modified cyclotron mode. First we couple the modified cyclotron mode to the axial detector by applying a sideband drive \cite{cornell1990mode} at $\nu_+-\nu_z$, this imprints the temperature of the axial mode $T_z$ to that of the cyclotron mode $T_+=(\nu_+/\nu_z)\cdot T_z$ \cite{brown1986geonium}, resulting in an initial thermal cyclotron radius 
\begin{eqnarray}
\rho_{+,\text{th}}=\sqrt{\frac{2k_B T_z}{m\omega_+^2}\frac{\nu_+}{\nu_z}}.
\end{eqnarray}
Subsequently we measure the axial frequency $\nu_{z,0}$ of the sideband-cooled particle and excite in a next step the modified cyclotron mode $\nu_+$ with a resonant drive that interacts with the particle for about $t_\text{exc}=700\,\mu$s. Projected to one dimension, this results in a particle radius
\begin{equation}
\label{eq:radiusincreaseresonantdrive}
\rho_{+,\text{exc}} =\left(\rho_{+,\text{th}}\cos(\phi_0)+\frac{q A_{\text{exc}}}{2 m \omega_{+}}t_\text{exc}\right),
\end{equation}
where $\phi_0$ is the particle's initial phase before the excitation and $A_{\text{exc}}$ the electrical field amplitude of the applied drive. The initial orbit radius $\rho_{+,\text{th}}$ adds incoherently to $\rho_{+,\text{exc}}$, and its contribution to the radius after excitation is invariant. After the excitation of the particle we measure $\nu_{z,\text{exc}}$ again and evaluate $\Delta\nu_z=\nu_{z,0}-\nu_{z,\text{exc}}$. By repeatedly applying this sequence we obtain the standard deviation  $\sigma(\Delta\nu_z)$.  Since $\Delta\nu_z\propto E_+$, dominantly determined by the fractional magnitude of the magnetic bottle strength $B_2/B_0$ \cite{ketter2014first}, the determination of $\sigma(\Delta\nu_z)$ is a direct measure of the cyclotron energy scatter
\begin{eqnarray}
\sigma(\Delta\nu_z)=\sqrt{2\Xi_\text{back}^2+\alpha_z\sigma(E_+)^2},
\end{eqnarray}
where $\Xi_\text{back}=26.2\,$mHz is the background scatter of the axial frequency measurements.
By calculating the standard deviation of $\sqrt{\langle E_+^2\rangle-\langle E_+\rangle^2}$, convolving $\rho_{+\text{exc}}$ with the initial radial thermal Rayleigh distribution after sideband-cooling, we obtain   
\begin{eqnarray}
\label{eq:excitationscatter}
\sigma(E_+)\approx \sqrt{2 E_{\text{th}}E_{\text{exc}}} \left(1+\frac{1}{4}\frac{E_{\text{th}}}{E_{+}}\right),
\end{eqnarray}
where $E_\text{th}=k_B (\nu_+/\nu_z)T_z$. The peak-technique thus continuously samples the axial temperature $T_z$, naturally implemented in the measurement campaign. Within $N$ measurements an axial temperature uncertainty of 
\begin{eqnarray}
\sigma(T_z)=\frac{\left(4\pi^2m\nu_z\right)^2}{\sqrt{2N-2}}\left(\frac{\nu_z}{\nu_+}\right)\left(\frac{B_0}{B_2}\right)^2\frac{\sigma(\Delta\nu_z^2)}{k_BE_+}
\end{eqnarray}
is obtained, we typically reach a 2.5$\,$K uncertainty within 50 samples. 
We verify this model by measuring $\sigma(E_+)$ as a function of the axial resonator temperature $T_z$, applying active feedback \cite{DUrso2003FeedbackOscillator} and by studying $\sigma(E_+)$ as a function of particle excitation energy. In addition, we measure the cyclotron frequency ratio of identical particles as a function of axial temperature and use as an additional consistency indicator the measured signal level difference $dS\propto 10\log((T_{\text{H}^-}\cdot Q_{\text{H}^-}\cdot\nu_{z,{\text{H}^-}})/(T_{\bar{\text{p}}}\cdot Q_{\bar{\text{p}}}\cdot\nu_{z,\bar{\text{p}}}))$. 
The determined axial temperatures, temperature differences and related frequency shifts are summarized in Tab$.\,$\ref{table:TEMP}

\subsection*{Pulling Shift}
In the peak method, the modified cyclotron frequency of the trapped particle is determined by exciting the modified cyclotron motion of the particle and subsequently measuring the frequency at which it deposits this excess energy into the cyclotron detector.
However, the frequency at which the particle is performing this damped oscillation is not purely determined by the Penning trap, but is also modified by the coupling to the detector, which imposes a dynamical image charge shift on the particle.
The resonance frequencies of this coupled system can be derived from the poles of the lineshape model in \eqref{Eq:Lineshape}.
When the ion damping is relatively small, as in case of the cyclotron detector, the resonance frequency of the damped particle can be approximated by
\begin{equation}
\nu_+= \nu_{+,0} +\frac{1}{4}\frac{1}{2 \pi \tau_+} \frac{  \Delta\nu_R (\nu_{+,0} - \nu_R)}{ (\nu_{+,0} - \nu_R)^2 +(\Delta\nu_R/2)^2}.
\end{equation}
For $\nu_{+,0}-\nu_R<0$ the measured cyclotron frequency is pulled downwards and opposite for $\nu_{+,0}-\nu_R>0$. In the peak campaign we adjust $\nu_R=\frac{1}{2}\left(\nu_{+,\bar{\text{p}}}+\nu_{+,{\text{H}^-}}\right)$. With $\nu_{+,\bar{\text{p}}}-\nu_{+,{\text{H}^-}}\approx32\,$kHz the measured frequency ratio $R_\text{exp}$ is shifted upwards by $\Delta R/R=2.86(24)\,$p.p.t. 

\subsection*{Summary of Frequency Shifts}
In this methods paragraph the dominant systematic frequency-ratio shifts were discussed, some additional suppressed frequency-ratio shifts are discussed in the supplementary material. All  considered frequency shifts are summarized in Tab$.\,$\ref{table:SUMM} which is displayed in the extended data figures. 

\section*{Standard Model Extension Coefficients}
The measurement of the antiproton-to-H$^-$ charge-to-mass ratio with a fractional precision of 16$\,$p.p.t$.$ enables us to provide improved constraints on coefficients of the standard-model extension (SME) \cite{kostelecky2011data}. A comprehensive manuscript discusses the impact of such measurements to searches for exotic physics, and gives a clear description on the derivation of CPT violating effects that couple to antiproton-to-H$^-$ charge-to-mass ratio comparisons \cite{ding2020lorentz}. From our experiment we derive the charge-to-mass ratio figure of merit 
\begin{eqnarray}
|\delta\omega_c^{\bar{p}}-R_{\bar{\text{p}},\text{p},\text{exp}}\delta\omega_c^{{p}}-2R_{\bar{\text{p}},\text{p},\text{exp}}\delta\omega_c^{{e}^-}|<1.96\cdot10^{-27}\,\text{GeV}, 
\end{eqnarray}
where $\frac{\delta\omega_c^{w}}{q_0 B}$ is a function of coefficients  $\tilde{b}_w$ and $\tilde{c}_w$ that describe the strengths of feebly interacting CPT-violating background fields, coupling to particles $w$, the antiproton $\bar{\text{p}}$, the proton p, and the electron $e^-$. By performing the transformation of the coefficients to the standard sun-centered frame \cite{kostelecky2011data}, following the theoretical outline given in \cite{ding2020lorentz} our measurement enables us to set improved limits on the coefficients summarized in Tab$.\,$\ref{table:SME}.

\section*{Acknowledgements} 

We acknowledge technical support by CERN, especially the Antiproton Decelerator operation group, CERN's cryolab team and engineering department, and all other CERN groups which provide support to Antiproton Decelerator experiments. We acknowledge Yunhua Ding for helpful comments in the discussion of the updated SME limits. 
We acknowledge financial support by RIKEN, the RIKEN EEE pioneering project funding, the RIKEN SPDR and JRA program, the Max-Planck Society, the European Union (FunI-832848,  STEP-852818), CRC 1227 "DQ-mat"(DFG 274200144), the Cluster of Excellence "Quantum Frontiers" (DFG 390837967), AVA-721559, the CERN fellowship program and the Helmholtz-Gemeinschaft. This work was supported by the Max-Planck, RIKEN, PTB-Center for Time, Constants, and Fundamental Symmetries (C-TCFS).

\section*{Author contributions statement}

The experiment was designed and built by S.U. and C.S., M.J.B., J.A.D., J.A.H., T.H. and E.J.W. developed several technical upgrades. J.A.H., S.U., T.H., J.A.D., E.J.W. and M.J.B. developed the control code. J.A.H., M.J.B., T.H., J.A.D., E.J.W. and S.U. took part in the data acquisition. M.J.B., S.U., J.A.D., J.A.H., E.J.W. and M.F. performed the systematic studies. J.A.H., M.J.B., T.H., J.A.D., E.J.W., S.R.E, and S.U. contributed to the maintenance of the experiment during the measurement campaign. The data were analyzed by S.U., E.W. and J.A.H., J.A.D., M.J.B., B.M.L., and C.W. contributed to the systematic analysis. The final results were discussed with all co-authors. The manuscript was written by S.U. and discussed with E.J.W., J.A.D, B.M.L., C.S. and K.B., all co-authors discussed and approved the content. 

\section*{Competing interests}
The authors declare no competing financial interests.


\section*{Data Records and Code Availability}

The data sets and analysis codes will be made available on reasonable request.
Correspondence and requests for materials should be addressed to ~\url{Stefan.Ulmer@cern.ch} .
\bibliography{sample}

\newpage

\section*{Extended Figures and Tables}

\begin{table}[h!]
\centering
\begin{tabular}{||c | c | c ||} 
\hline
Campaign & Correction & Uncertainty  \\ [0.5ex] 
\hline\hline
2018-1-SB & $-0.37\,$ p.p.t$.$ & $20.65\,$p.p.t$.$   \\ 
2018-2-SB & $-16.89\,$ p.p.t$.$ & $46.49\,$p.p.t$.$   \\
2018-3-PK & $-0.74\,$ p.p.t$.$ & $0.61\,$ p.p.t$.$   \\
2019-1-SB & $8.61\,$ p.p.t$.$ & $21.45\,$p.p.t$.$   \\
\hline
\end{tabular}
\caption{Summary of lineshape-corrections applied to the different data sets}
\label{table:CORR}
\end{table}

\begin{table}[h!]
\centering
\begin{tabular}{||c | c | c | c | c | c ||} 
\hline
Run & $T_{z,\text{H}}$ (K) & $T_{z,\text{p}}$ (K)& $\Delta T_z$ (K) & $B_2$-shift (p.p.t.)& rel. shift (p.p.t$.$)   \\ [0.5ex] 
\hline\hline
18-1-SB & $6.27(14)\,$  & $5.98(15)\,$ & 0.29(21) & $20.27(14.86)\,$ & 1.20(92)   \\ 
18-2-SB & $6.16(14)\,$  & $6.04(15)\,$ & 0.12(21)  & $8.38(14.86)\,$ & 0.47(90)   \\
18-3-PK & $11.31(40)\,$  & $10.85(35)\,$ & 0.46(56) & $10.79(12.66)\,$ & 1.90(2.32)  \\
19-1-SB & $5.57(16)\,$  & $5.41(15)\,$ & 0.16(22) & $3.75(5.16)\,$ & 0.65(94)   \\
\hline
\end{tabular}
\caption{Summary of measured axial temperatures, axial temperature differences and ratio corrections of the different measurement campaigns. }
\label{table:TEMP}
\end{table}

\begin{table*}[h!]
\centering
\begin{tabular}{||l | c | c | c | c ||} 
\hline
Effect & 2018-1-SB  & 2018-2-SB & 2018-3-PK & 2019-1-SB   \\ [0.5ex] 
\hline\hline
&&&&
\\
$B_1$-shift & $0.03(2)$ & $0.01(2)$ & $<(0.01)$ & $<(0.01)$   
\\ 
$B_2$-shift & $20.27(14.86)$ & $8.38(14.86) $ & $10.79(12.66)$  & 3.75 (5.16)   
\\
$C_4$-shift & (1.12) & $(1.13)$ & $(1.54)$ & (0.76)
\\
$C_6$-shift & $<(0.01)$ & $<(0.01)$ & $<(0.01)$ & $<(0.01)$ 
\\
Relativistic & $1.20(92)$ & $0.47(90)$ & $1.90(2.32)$ & $0.65(94)$ 
\\
&&&&
\\
Image charge shift & $0.05(0)$  & $0.05(0)$ & $0.05(0)$ & $0.05(0)$   
\\
Trap misalignment & $0.06(0)$  & $0.06(0)$ & $0.05(0)$ & $0.05(0)$  
\\
&&&&\\
Voltage Drifts & $-3.35(5.12)$ & $-3.77(5.12)$ & $-0.11(11)$ & $-5.03(5.12)$
\\
Spectrum Shift & $0.37(20.65)$ & $16.89(46.49)$ & $0.74(61)$ & $-8.61 (21.45)$
\\
FFT-Distortions & (1.57) & $(3.48)$ & $(0.03)$ & $(1.23)$
\\
Resonator-Shape & $0.02(3)$ & $0.02(2)$ & $<(0.01)$ & $0.01(2)$
\\
&&&&
\\
$B_1$-drift offset & $<(0.11)$ & $<(0.11)$ & $<(0.04)$ & $<(0.04)$
\\
Resonator Tuning & $<(0.16)$ & $<(0.16)$ & $<(0.06)$ & $<(0.06)$
\\
&&&&
\\
Averaging Time & $-$ & $-$ & $-2.87(25)$ & $-$
\\
FFT Clock & $-$ & $-$ & $(3.69)$ & $-$
\\
Pulling Shift & $-$ & $-$ & $2.86(24)$ & $-$
\\
Linear Coefficient Shift& $-$ & $-$ & $0.16(40)$ & $-$
\\
Nonlinear Shift & $-$ & $-$ & $0.03(2)$ & $-$
\\
\hline
\hline
&&&&
\\
Systematic Shift & $18.65(26.04)$ & $22.11 (49.22)$ & $13.60(13.50)$ & $-9.13(22.71)$
\\
&&&&\\
\hline
&&&&\\
$R_{\text{exp}}-R_{\text{theo}}$ & $13.02(27.12)$ & $-5.04(46.57)$ &$ 7.99(18.57) $ & $18.34(18.89)$\\
&&&&\\
\hline
\hline
&&&&\\
$R_{\text{exp,c}}-R_{\text{theo}}$ & $-5.63(37.60)$ & $-27.15(67.76)$ &$ -5.61(22.66) $ & $27.47(29.54)$\\
&&&&\\
\hline
\end{tabular}
\caption{Summary of systematic shifts and uncertainties for the sideband (SB) and peak (PK) campaigns. The peak measurement method suppresses the dominant systematic contribution of the SB method by a factor of $\nu_z/\nu_+$. Table entries are in p.p.t$.$ units. }
\label{table:SUMM}
\end{table*}

\begin{table*}[h!]
\centering
\begin{tabular}{||l | c | c | c | c ||} 
\hline
Coefficient  &Previous Limit & Improved Limit & Factor  \\ [0.5ex] 
\hline\hline
$|\tilde{c}_e^{XX}|$  & $<3.23\cdot10^{-14}$ & $<7.79\cdot10^{-15}$ & 4.14  \\
$|\tilde{c}_e^{YY}|$  & $<3.23\cdot10^{-14}$ & $<7.79\cdot10^{-15}$ & 4.14  \\
$|\tilde{c}_e^{ZZ}|$  & $<2.14\cdot10^{-14}$ & $<4.96\cdot10^{-15}$ & 4.31  \\
\hline
\hline
$|\tilde{c}_p^{XX}|, |\tilde{c}_p^{*XX}|$  & $<1.19\cdot10^{-10}$ & $<2.86\cdot10^{-11}$ & 4.14  \\
$|\tilde{c}_p^{YY}|, |\tilde{c}_p^{*YY}|$  & $<1.19\cdot10^{-10}$ & $<2.86\cdot10^{-11}$ & 4.14  \\
$|\tilde{c}_p^{ZZ}|, |\tilde{c}_p^{*ZZ}|$  & $<7.85\cdot10^{-11}$ & $<1.82\cdot10^{-11}$ & 4.31  \\
[1ex] 
\hline
\end{tabular}
\caption{Constraints on coefficients of the standard model extension. The second column describes the previous best limit based on \cite{gabrielse1999precision} and \cite{Ulmer2015High-precisionRatio},  theorized and summarized in \cite{ding2020lorentz}. The third column gives the improved limit based on the measurement presented here, the fourth column shows the ratio of the fourth and the third column. All entries are based on C.L$.$0.68. }
\label{table:SME}
\end{table*}

\pagebreak

\end{document}